\documentclass[twocolumn,showpacs,preprintnumbers,amsmath,amssymb]{revtex4}

\usepackage{graphicx}
\usepackage{dcolumn}
\usepackage{bm}
\usepackage{amssymb}

\def\address{\@ifstar{\address@star}%
  {\@ifnextchar[{\address@optarg}{\address@noptarg}}}

\begin{document}

\author{S.N.~Gninenko$^{1}$}
\author{N.V.~Krasnikov$^{1}$}
\author{A.~Rubbia$^{2}$}

\affiliation{$^{1}$Institute for Nuclear Research of the Russian Academy of Sciences, Moscow 117312}
\affiliation{$^{2}$Institut f\"ur Teilchenphysik, ETHZ, CH-8093 Z\"urich, Switzerland}


\title{Extra dimensions and invisible decay of orthopositronium}

\date{\today}

\begin{abstract} 
We  point out that some  models with infinite 
additional dimension(s) of Randall-Sundrum type predict the disappearance of 
orthopositronium ($o-Ps$) into additional  dimension(s). 
The experimental signature of this effect is the  
$o-Ps \to invisible$ decay of orthopositronium which  may  occur at a rate 
within  three orders of magnitude of the present experimental upper 
limit. This result enhances  existing motivations for a 
more sensitive search for this decay mode and suggests additional directions 
for testing  extra dimensions in non accelerator experiments. 
\end{abstract}

\pacs{14.80.-j, 12.60.-i, 13.20.Cz, 13.35.Hb}
\maketitle

Positronium ($Ps$), the positron-electron bound state,
 is the lightest known atom, which is bounded and self-annihilates 
through the same, electromagnetic interaction. At the 
current levels of experimental and theoretical precision this is  
the only interaction present in this system, see e.g. \cite{rich}. 
This feature has made positronium an ideal system for  
testing of the QED calculations accuracy
for bound states, in particular for the triplet ($1^3S_1$)
state of $Ps$, orthopositronium ($o-Ps$). 
Due to the odd-parity under
C-transformation  $o-Ps$ decays
predominantly into three photons. 
 As compared with singlet ($1^1S_0$) state (parapositronium),
 the "slowness" of $o-Ps$ decay, due to the phase-space and  additional
$\alpha$ suppression factors, gives an enhancement factor $\simeq 10^3$,
making it more sensitive to an admixture of 
new interactions which are not accommodated in the Standard Model (SM), 
see e.g. \cite{do}.

Within the SM orthopositronium can  decay invisibly into
 neutrino-antineutrino pair.  
The $o-Ps \to \nu_e \bar{\nu}_e$ decay occurs through 
$W$ exchange in $t$ channel and $e^+e^-$ annihilation via $Z$.
The decay width is \cite{czar}
\begin{eqnarray} 
\Gamma(o-Ps \to \nu_e  \bar{\nu}_e) 
\approx 6.2 \cdot 10^{-18}\Gamma(o-Ps \to 3\gamma)
\end{eqnarray}
For other neutrino flavours only the $Z$-diagram 
contributes. For $ l \neq e$ the decay width is \cite{czar}
\begin{eqnarray} 
\Gamma(o-Ps \rightarrow \nu_l \bar{\nu}_l) 
\approx 9.5 \cdot 10^{-21}\Gamma(o-Ps \to 3\gamma)
\end{eqnarray}
Thus, in the SM the $o-Ps \to \nu \bar{\nu}$ decay width
is very small and its contribution to the total decay width can  
be neglected. 

Presently there is a big interest in models with
additional dimensions which might provide  solution to the 
gauge hierarchy problem \cite{randall}-\cite{giud},
 for a recent review see e.g. \cite{rubakov}.
For instance, as it has been shown  
in the five dimensional model, so called RS 2-model \cite{randall},  
there exists a thin-brane solution to the 5-dimensional 
Einstein equations which has flat 4-dimensional hypersurfaces,
\begin{equation}
ds^2 = a^2(z)\eta_{\mu\nu}dx^{\mu}dx^{\nu} -dz^2.
\end{equation} 
Here
\begin{equation}
a(z) = exp(-k|z|)
\end{equation}
and the parameter $k > 0 $ is determined by the 5-dimensional Planck 
mass and bulk cosmological constant.
 
Recently, a peculiar feature of massive matter in brane world has been 
reported \cite{tinyakov}. It has been shown that tunnelling of massive matter 
into extra dimensions is a generic to fields that can have 
bulk modes. The massive matter becomes 
unstable, namely the discrete zero modes turn into quasi-localised 
states with finite 4-dimensional mass and finite width \cite{tinyakov}.
For massive scalar particle $\Phi$ with the mass $m$ the transition rate 
into additional dimension is given by \cite{tinyakov}
\begin{equation}
\Gamma(\Phi \rightarrow add.~dim.) =
\frac{\pi m}{16}(\frac{m}{k})^2
\end{equation}

It should be noted that even for a massless scalar particle the 
nonzero transition rate into additional dimension(s) results in 
a nonzero imaginary part of the corresponding scalar propagator 
\cite{tinyakov} 
\begin{equation}
D(p^2) = \frac{1}{p^2 - i\theta(p^2)\sqrt{p^2}\Gamma(p^2)}
\end{equation}
where $\Gamma(p^2) \approx\frac{\pi}{4} \sqrt{p^2}(\frac{p^2}{k^2})$.
 It means that 
even for the massless case, when the transition rate on mass shell 
is zero, the virtual scalar 
particle has nonzero transition rate into additional dimension. 
The explicit expression for the transition rate into additional dimension 
depends on the concrete model. 

To be specific, let us consider a model for the localisation of gauge 
fields suggested in ref.\cite{oda,rubilnik}. One begins with the solution
to $(4+n+1)$-dimensional Einstein equations,

\begin{equation}
ds^2=\frac{1}{(1+k|\xi | )^2}\bigl(dt^2-d{\bf x}^2-\sum_{i=1}^n R^2_i d\theta_i^2 - d\xi^2\bigr)
\end{equation}
where $\theta_i \in [0,2\pi]$ are compact coordinates, $R_i$ are radii of 
compact dimensions and $k$ is the inverse adS radius determined by the bulk 
cosmological constant. There is a single brane located at $\xi = 0$. The only
difference between this metric and the Randall-Sundrum metric is the 
presence of extra compact dimensions $\theta_i$. These dimensions are added 
for obtaining a localised zero mode of the gauge field. In what follows we 
assume that their radii $R_i$ are the smallest length scales involved, so 
all fields are taken independent of $\theta_i$. The inverse adS radii $k$
is assumed to be the largest energy scale involved. For the model of 
ref.\cite{oda,rubilnik} with additional (n+1) dimensions and metric of 
Randall-Sundrum type the imaginary part  of the propagator 
of massless scalar particle  is 
$\Gamma \sim \sqrt{p^2}(\frac{p^2}{k^2})^{1+n/2}$. 

 The case of the electromagnetic field propagating in 
the Randall-Sundram type of metric   of Eq.(7)
has been considered in ref.\cite{rubilnik}. It was shown that 
 the transition rate of a virtual photon with the 
virtual mass $m_{\gamma^{*}} =\sqrt{p^2}$ into additional dimensions is 
different from zero and it is equal to

\begin{equation} 
\Gamma = k(n) m_{\gamma^{*}}(\frac{m_{\gamma^{*}}}{k})^{n} 
\end{equation} 
where $k(n)$ a numerical coefficient.

Consider now   $o-Ps \to invisible$ decay, which is  a good candidate 
for the searching for effect of disappearance into 
additional dimension(s) since $o-Ps$  has specific quantum numbers similar 
to those of vacuum and is a system which allows its constituents a rather 
long interaction time.
To make quantitative estimate  we take $n=2$ ( (4+2+1)-dimensional
space-time). In this case  disappearance rate  of a virtual
 photon  into additional dimensions is given by 
\begin{equation} 
\Gamma (\gamma^* \to add. ~dim.)=\frac{\pi m_{\gamma^{*}}}{4}\Bigl(\frac{m_{\gamma^{*}}}{k}\Bigr)^2  
\end{equation}    

Using Eq.(9), for the branching ratio of orthopositronium invisible decay 
into  additional dimension(s) through single photon annihilation 
$o-Ps \rightarrow \gamma^{*} \rightarrow  add.~dim.$ 
 one gets as an estimate
\begin{eqnarray}
&&\frac{\Gamma(o-Ps \to  \gamma^{*} \rightarrow add.~dim.)}
{\Gamma(o-Ps \to  3\gamma)}= \frac{9\pi}{4(\pi^2-9)}
\frac{1}{\alpha^2}\cdot \nonumber \\
&&\frac{\pi}{4}(\frac{m_{o-Ps}}{k})^2 \approx 1.2\cdot10^{5}(\frac{m_{o-Ps}}{k})^2
\end{eqnarray}

To solve the gauge hierarchy 
problem models with additional dimension(s) one may expect
$k \lesssim O(10)~TeV$. It means that   
\begin{equation}
Br(o-Ps \rightarrow add.~dim.) \gtrsim O(10^{-9})
\end{equation}
Important bounds on the parameter $k$ and 
$Br(o-Ps \rightarrow add.~dim.)$ arise from the combined LEP result
 on the precise measurements of the total and partial $Z$ widths \cite{pdg}.
We can write the $Z$ invisible  width in the following form:
\begin{equation}
\Gamma_{inv} = \Gamma^{SM}(Z \to \nu \overline{\nu})+\Delta \Gamma_{inv}
\end{equation} 
where $\Gamma^{SM}(Z \to \nu \overline{\nu})$ is the SM 
contribution and $\Delta \Gamma_{inv}$ contains the effects beyond the SM.
Assuming that each neutrino type contributes the same amount to the 
invisible $Z$ width, one has numerically \cite{peccei}
\begin{eqnarray}
&&\Gamma^{SM}(Z \to \nu \overline{\nu})=
3\Gamma(Z \to \nu_i \overline{\nu}_i) \nonumber \\
&&=3\cdot (167.06\pm 0.22)~MeV 
\end{eqnarray} 
The invisible width $\Gamma_{inv}$ can  be obtained from
 the $Z$ total width and  its partial width
into hadrons and leptons using the equation
\begin{equation}
\Gamma_{tot} = \Gamma_{had} +\Gamma_{lept}+\Gamma_{inv}
\end{equation} 
The value of $\Gamma_{inv}$ derived from the LEP measurements of   
$\Gamma_{tot},~\Gamma_{had}$ and $\Gamma_{lept}$  
\cite{pdg,lep} is 
\begin{equation}
\Gamma_{inv} = 499.0\pm1.5~MeV
\end{equation} 
Using (13) and (15) we obtain
\begin{equation}
\Delta \Gamma_{inv} = -2.7\pm 1.6 ~MeV
\end{equation} 
If a conservative approach is taken to constrain the result to only positive 
values $\Delta \Gamma_{inv}$ and renormalising the probability for 
$\Delta \Gamma_{inv}\geq 0$ to be unity, then the resulting 
 95\% CL upper limit on additional invisible decay of $Z$ is \cite{lep} 
\begin{equation}
\Delta \Gamma_{inv} < 2.0 ~ MeV 
\end{equation}
Assuming $\Delta \Gamma_{inv}=\Gamma(Z\to add.~dim.)$ and using  Eqs.(9,10) 
for the estimate leads to $k \geq 17~TeV$ 
and to the corresponding bound:
\begin{eqnarray}
Br(o-Ps \rightarrow add.~dim.) \leq  0.4\cdot 10^{-9}
\end{eqnarray}
Note that combined result on direct LEP measurements of the invisible width,
$\Gamma_{inv}=503\pm 16 ~MeV$ \cite{pdg} gives less stringent limit   
\begin{eqnarray}
Br(o-Ps \rightarrow add.~dim.) \lesssim  10^{-8}
\end{eqnarray}

These estimates  giving only an order of magnitude for the
 corresponding branching ratio show that this decay 
 may occur at a rate within roughly three
 orders of magnitude of the best present experimental 
limit \cite{mits}: 
\begin{equation}
Br(o-Ps \rightarrow invisible ) < 2.8 \cdot 10^{-6} 
\end{equation}
Thus, the region $Br(o-Ps\to invisible) \simeq 10^{-9}$ is of
 great interest for possible  observation of effect of extra dimensions. 
Interestingly, that  for $n=1$ the bound 
$Br(o-Ps \rightarrow add.~dim.) \lesssim  10^{-4}$ obtained  
from (17) is weaker than that of (20) obtained  from the direct measurement. 
We believe these results strengthens  current motivations related to the 
orthopositronium decay rate puzzle and  mirror world \cite{glashow}, 
 millicharged particle \cite{prinz} and light gauge boson \cite{do} searches, 
and justify efforts 
for a more sensitive search for the $o-Ps\to invisible$ decay 
in a near future experiment \cite{ethz}. 

The experimental signature of  the $o-Ps\rightarrow invisible$ decay  
is the absence of an energy deposition of $\simeq$ 1 MeV,
which is expected from the ordinary $o-Ps$ annihilation
in a 4$\pi$ hermetic calorimeter surrounding the $o-Ps$ formation region
\cite{atojan}.
Our first Monte Carlo simulations, based on the results of the recent 
search for  $o-Ps\to \gamma + invisible$ decay \cite{bader},
show that for the branching ratio one may expect a limit 
 $Br(o-Ps \rightarrow add.~dim.) \lesssim10^{-8}$ 
if the calorimeter has a mass of $\simeq$ 0.5 ton. Larger  simulation 
statistics and better background evaluation are required 
in order to see if the sensitivity to the branching ratio as low as $10^{-9}$
is experimentally reachable.
  
We would like to thank a referee for useful remarks. 
N.V.K. is indebted to V.A. Rubakov and P.G. Tinyakov
 for many useful discussions on models with additional infinite dimensions.
The work of N.V.K. has been supported by the RFFI Grant N02-01-00601.

\end{document}